\documentstyle[12pt]{article}

\input{epsf.sty}

\def\plotfiddle#1#2#3#4#5#6#7{\centering \leavevmode
    \vbox to#2{\rule{0pt}{#2}}
    \includegraphics{#1}}

\begin{document}

\large

\centerline{{\huge \sc Groups of Galaxies in the}}
\vspace{0.50cm}
\centerline{{\huge \sc Las Campanas Redshift Survey}}
\vspace{1.50cm}
\centerline{{\large Douglas L. Tucker (Fermilab)}}
\vspace{0.50cm}
\centerline{{\large Yasuhiro Hashimoto (Yale)}}
\vspace{0.50cm}
\centerline{{\large Robert P. Kirshner (CfA)}}
\vspace{0.50cm}
\centerline{{\large Stephen D. Landy (UC/Berkeley)}}
\vspace{0.50cm}
\centerline{{\large Huan Lin (Toronto)}}
\vspace{0.50cm}
\centerline{{\large Augustus Oemler, Jr. (OCIW)}}
\vspace{0.50cm}
\centerline{{\large Paul L. Schechter (MIT)}}
\vspace{0.50cm}
\centerline{{\large Stephen A. Shectman (OCIW)}}
\vspace{3.50cm}

\begin{abstract}
A ``friends-of-friends'' percolation algorithm has been used to
extract a catalogue of $\delta\rho/\rho = 80$ density enhancements
(groups) from the six slices of the Las Campanas Redshift Survey
(LCRS).  The full catalogue contains 1495 groups and includes 35\% of
the LCRS galaxy sample.  A statistical sample of 394 groups has been
derived by culling groups from the full sample which either are too
close to a slice edge, have a crossing time greater than a Hubble
time, have a corrected velocity dispersion of zero or less, or contain
a 55~arcsec ``orphan'' (a galaxy with a ``faked'' redshift excluded
from the original LCRS redshift catalogue due to its proximity ---
i.e., within 55~arcsec --- of another galaxy).  Median properties
derived from the statistical sample include: line-of-sight velocity
dispersion $\sigma_{\rm los} = 164$~km~s$^{-1}$, crossing time $t_{\rm
cr} = 0.10~H_0^{-1}$, harmonic radius $R_{\rm h} = 0.58~h^{-1}$~Mpc,
pairwise separation $R_{\rm p} = 0.64~h^{-1}$~Mpc, virial mass $M_{\rm
vir} = 1.90~\times~10^{13}~h^{-1}~M_{\rm sun}$, total group $R$-band
luminosity $L_{\rm tot} = 1.40~\times~10^{11}~h^{-2}~L_{\rm sun}$, and
$R$-band mass-to-light ratio $M/L = 153~h~M_{\rm sun}/L_{\rm sun}$.
\end{abstract}

\clearpage

\section{The Las Campanas Redshift Survey (LCRS)}

The Las Campanas Redshift Survey (LCRS; Shectman et al.\ 1996) is an
optically selected galaxy redshift survey which extends to a redshift
of 0.2 and which is composed of a total of 6 alternating $1.5^{\circ}
\times 80^{\circ}$ slices, 3 each in the North and South Galactic
Caps.  Now completed, the LCRS contains 26,418 galaxy redshifts, of
which 23,697 lie within the official geometric and photometric limits
of the survey.  Accurate $R$-band photometry and sky positions for
program objects were extracted from CCD drift scans obtained on the
Las Campanas Swope 1-m telescope; spectroscopy was performed at the
Las Campanas Du Pont 2.5-m telescope, originally via a 50-fiber
Multi-Object Spectrograph (MOS), and later via a 112-fiber MOS.  For
observing efficiency, all the fibers were used, but each MOS field was
observed only once.  Hence, the LCRS is a collection of 50-fiber
fields (with nominal apparent magnitude limits of $16.0 \leq R <
17.3$) and 112-fiber fields (with nominal apparent magnitude limits of
$15.0 \leq R < 17.7$); see {\bf Figure~1}.  Thus, selection criteria
vary from field to field, but these selection criteria are carefully
documented and therefore easily taken into account.  Observing each
field only once, however, created an additional selection effect: the
protective tubing of the individual fibers prevented the spectroscopic
observation of both members of galaxy pairs within 55~arcsec of each
other.  Hence, groups and clusters can be undersampled, potentially
causing physical groups to be split by a ``friends-of-friends''
percolation algorithm and resulting in the mis-estimate of general
group properties.  We will return to this problem in the next section.

\section{Extracting The Group Catalogue}

In constructing the LCRS Group Catalogue, we have considered only
those LCRS galaxies within the official geometric and photometric
borders of the survey; we have furthermore limited this sample to
galaxies having redshifts in the range
\begin{equation}
7,500~\mbox{km~s}^{-1} \leq cz_{\rm cmb} < 50,000~\mbox{km~s}^{-1}
\end{equation}
and luminosities in the range
\begin{equation}
-22.5 \leq M_R - 5\log h < 17.5
\end{equation}
(see {\bf Figure~2}).  

Moreover, each of the $\sim 1,000$ galaxies which were excluded from
LCRS redshift catalogue due to the fiber-separation effect has been
re-introduced into the sample by assigning it a redshift equal to the
redshift of its nearest neighbor convolved with a gaussian of width
$\sigma = 200$~km~s$^{-1}$ (roughly the mean line-of-sight velocity
dispersion of a cleaned LCRS group sample which excludes these
55-arcsec ``orphans'').  The re-included galaxies subscribe to all the
same limits imposed upon the original sample.

The group catalogue was extracted using a standard
``friends-of-friends'' percolation algorithm (Huchra \& Geller 1982)
modified for comoving distances and for field-to-field sampling
variations.  To take into account the latter, the projected separation
and velocity difference linking parameters, $D_L$ and $V_L$, respectively, 
were scaled according to the following equations (which assure that
the ratio $D_L/V_L$ is independent of environment):
\begin{equation}
D_L = D_0 \times S_L \hspace{1.0cm} \mbox{and} \hspace{1.0cm} V_L = V_0 \times S_L, 
\end{equation}
where $D_0$ and $V_0$ are $D_L$ and $V_L$, respectively, for a given
fiducial field at at given fiducial redshift, and where $S_L$ is a
linking scale which takes into account variations in galaxy sampling
rate.  It is defined by
\begin{equation}
S_L \equiv \left[ \frac{ \rho^{\rm exp}(f,z) }{ \rho^{\rm exp}_{\rm fid}} \right]^{-1/3}
\end{equation}
where $\rho^{\rm exp}(f,z)$ is the number density of galaxies one
would expect to observe at redshift $z$ in field $f$ for a homogeneous
sample having the same selection function and sampling fraction as the
LCRS redshift catalogue; $\rho^{\rm exp}_{\rm fid}$ is $\rho^{\rm
exp}(f,z)$ for a given fiducial field at a given fiducial redshift.
Due to the simple field characteristics, we have chosen the fiducial
field to have 100\% sampling, flux limits of $15.0 \le R < 17.7$, and
the same luminosity function as the LCRS Northern 112-fiber sample
(Lin et al.\ 1996); since it is roughly the median redshift of the
survey, we have chosen the fiducial redshift $cz_{fid}$ to be
30,000~km~s$^{-1}$.

Finally, to avoid group-member incompleteness at the extremal
distances of the sample, only groups within
\begin{equation}
10,000~\mbox{km~s}^{-1} \leq cz_{\rm cmb} < 45,000~\mbox{km~s}^{-1}
\end{equation}
were admitted into the final group catalogue.

\section{Properties of LCRS Groups}

The full catalogue contains 1495 groups and includes 35\% of the LCRS
galaxy sample ({\bf Figures~3} and {\bf 4}).  A statistical sample of
394 groups was extracted from the full sample by culling groups which
either were too close to a slice edge, had a crossing time greater
than a Hubble time, had a corrected velocity dispersion of zero or
less, or contained a 55~arcsec ``orphan''.  Some of the
characteristics of the statistical sample are listed in {\bf Table I},
including medians of the following group properties:

\begin{itemize}

\item 
The group line-of-sight velocity dispersion, $\sigma_{\rm los}$,
corrected for relativistic effects (Harrision 1974) and for estimated
random errors in the LCRS redshifts.

\item 
The mean pairwise separation,
\begin{equation}
R_{\rm p} = \frac{8 D_{\rm grp}}{\pi} 
\sin \left\{ \frac{1}{2} 
\left[ \frac{\sum_{i} \sum_{j>i} w_i w_j \theta_{ij}}{\sum_{i} \sum_{j>i} w_i w_j} \right]
\right\} ,
\end{equation}
where $D_{\rm grp}$ is the comoving distance to the group,
$\theta_{ij}$ is the angular separation between group members $i$ and
$j$, and $w_i$ and $w_j$ are the respective weights for
$i$ and $j$,
\begin{equation}
w_i \equiv \left[ \rho^{\rm exp}(f_i,z_i) \right]^{-1}
\end{equation}
This weighting factor helps to counteract a bias resulting from a
group straddling two fields with different galaxy sampling
characteristics.

\item 
The harmonic radius,
\begin{equation}
R_{\rm h} = \pi D_{\rm grp} 
\sin \left\{ \frac{1}{2} 
\left[ \frac{\sum_{i} \sum_{j>i} w_i w_j \theta_{ij}^{-1}}{\sum_{i} \sum_{j>i} w_i w_j} \right]^{-1} 
\right\} .
\end{equation}

\item 
The crossing time for the group,
\begin{equation}
t_{\rm cr} = \frac{3}{5^{3/2}} \frac{R_{\rm h}}{\sigma_{\rm los}} ,
\end{equation}
in units of the Hubble time ($H_0^{-1}$).  

\item 
The group's virial mass,
\begin{equation}
M_{\rm vir} = \frac{6 \sigma_{\rm los}^2 R_{\rm h}}{G} ,
\end{equation}
where $G$ is the gravitational constant.

\item
The total group luminosity in the LCRS $R$-band, $L_{\rm tot}$,
corrected via the selection function to account for galaxies not
observed by the LCRS.

\item
The group mass-to-light ratio in the LCRS $R$-band, $M/L$.

\end{itemize}
The above definitions are very similar to those used by Ramella,
Geller, \& Huchra (1989), but modified to take into account
cosmological effects (due to the LCRS sample depth) and field-to-field
sampling variations.

{\bf Table I} lists the properties of LCRS groups from the whole
statistical sample, of those groups from just the 50-fiber fields, of
those groups from just the 112-fiber fields, and of those groups which
straddle the border of a 50-fiber and a 112-fiber field; also
tabulated are the general properties from an earlier incarnation of
the LCRS $\delta=-6^{\circ}$ group catalogue [Tucker 1994 (T94)].  The
50-fiber-field groups appear to be typically a little larger than the
112-fiber-field groups in both projected size ($<R_{\rm p}>_{\rm med}$
\& $<R_{\rm h}>_{\rm med}$) and velocity extent ($<\sigma_{\rm
los}>_{\rm med}$), indicating that the {\em effective} linking scale
for the 50-fiber fields may be systematically larger than that for the
112-fiber fields, in spite of efforts to avoid such a bias.  On the
other hand, much of the apparent divergence of the 50/112 group
properties can be attributed to this sample's aberrantly high median
velocity dispersion (e.g., recall $M_{\rm vir} \propto \sigma_{\rm
los}^2$), which in turn may be due to the small size of the 50/112
statistical sample or to the difficulties of properly extracting
groups which straddle the border between a 50-fiber and a 112-fiber
field.

Finally, for comparison, we have also listed in {\bf Table I} the
median group properties from several other group catalogues.  These
include those based upon the the original CfA redshift survey [CfA1;
Nolthenius \& White 1987 (NW87); Noltenius 1993 (N93); Moore, Frenk,
\& White 1993 (MFW93)], the Southern Sky Redshift Survey [SSRS; Maia,
da~Costa, \& Latham 1989 (MdCL89)], and the CfA extension to $m_{\rm
B(0)} = 15.5$ [CfA2; Ramella, Geller, \& Huchra 1989 (RGH89); Ramella,
Pisani, \& Geller 1997 (RPG97)].

\clearpage
 
\section*{References}

\begin{description}

\item[]
Harrison, E.~R. 1974, ApJ, 191, L51

\item[]
Huchra, J.~P., and Geller, M.~J. 1982, ApJ, 257, 423

\item[] 
Lin, H., Kirshner, R.~P., Shectman, S.~A., Landy, S.~D., Oemler, A.,
Tucker, D.~L., and Schechter, P.~L. 1996, ApJ, 464, 60

\item[]
Maia, M.~A.~G., da~Costa, L.~N., and Latham, D.~W. 1989, ApJS,
69, 809 (MdCL89)

\item[]
Moore, B., Frenk, C.~S., and White, S.~D.~M. 1993, MNRAS, 261, 827 (MFW93)

\item[]
Nolthenius, R., and White, S.~D.~M. 1987, MNRAS, 225, 505 (NW87)

\item[]
Nolthenius, R. 1993, ApJS, 85, 1 (N93)

\item[]
Ramella, M., Geller, M.~J., and Huchra, J.~P. 1989, ApJ, 344, 57 (RGH89)

\item[]
Ramella, M., Pisani, A., Geller, M.~J. 1997, AJ, 113, 483 (RGP97)

\item[]
Shectman, S.~A., Landy, S.~D., Oemler, A., Tucker, D.~L., Lin, H.,
Kirshner, R.~P., and Schechter, P.~L. 1996, ApJ, 470, 172

\item[]
Tucker, D.~L. 1994, Ph.D. dissertation, Yale University (T94)

\end{description}

\section*{Acknowledgments}
This research has made use of the NASA/IPAC Extragalactic Database
(NED), which is operated by the Jet Propulsion Laboratory, Caltech,
under contract with the National Aeronautics and Space Administration.

\clearpage

\begin{figure}
\plotfiddle{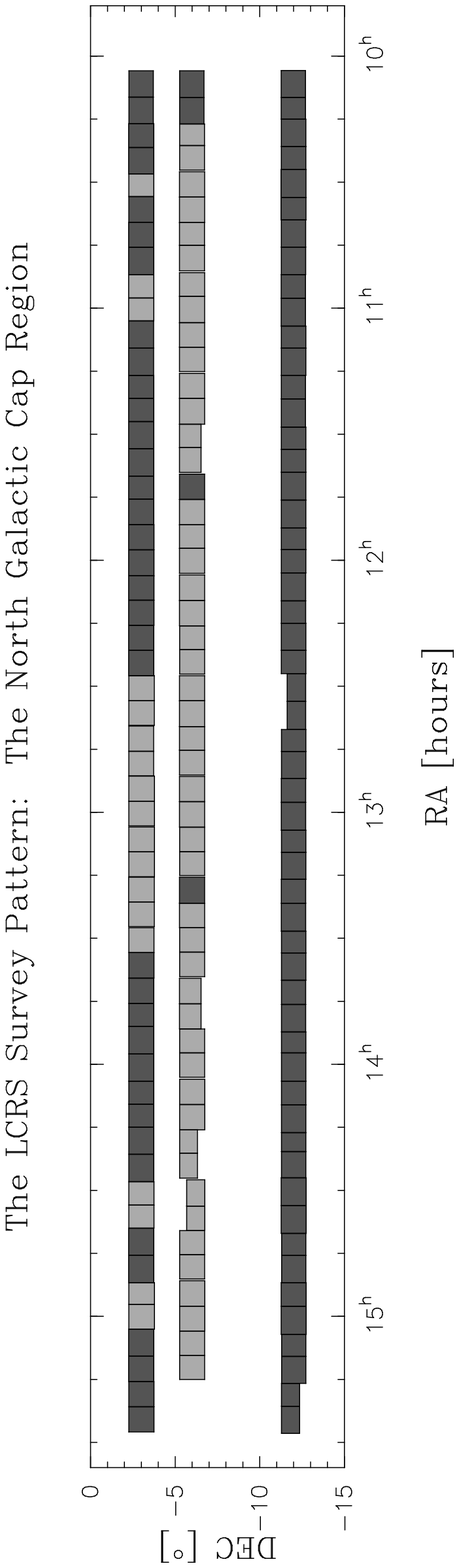}{5.00in}{-90}{70}{70}{-275}{525}
\plotfiddle{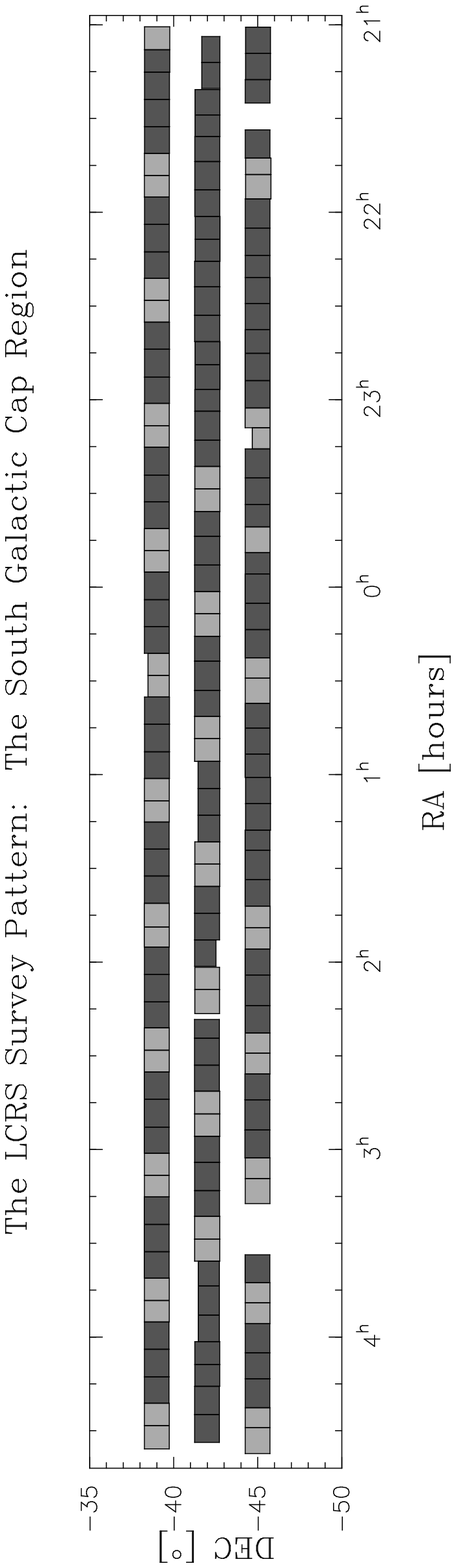}{5.00in}{-90}{70}{70}{-275}{690}
\vspace{-10.0cm}
\caption{The LCRS survey pattern for the Northern (top) and the Southern (bottom) Galactic
Cap regions.  Lightly shaded regions denote fields observed with the 50-fiber MOS and
darkly shaded regions fields observed with the 112-fiber MOS.  Declination and right
ascension coordinates are epoch 1950.0}
\end{figure}

\clearpage

\begin{figure}
\plotfiddle{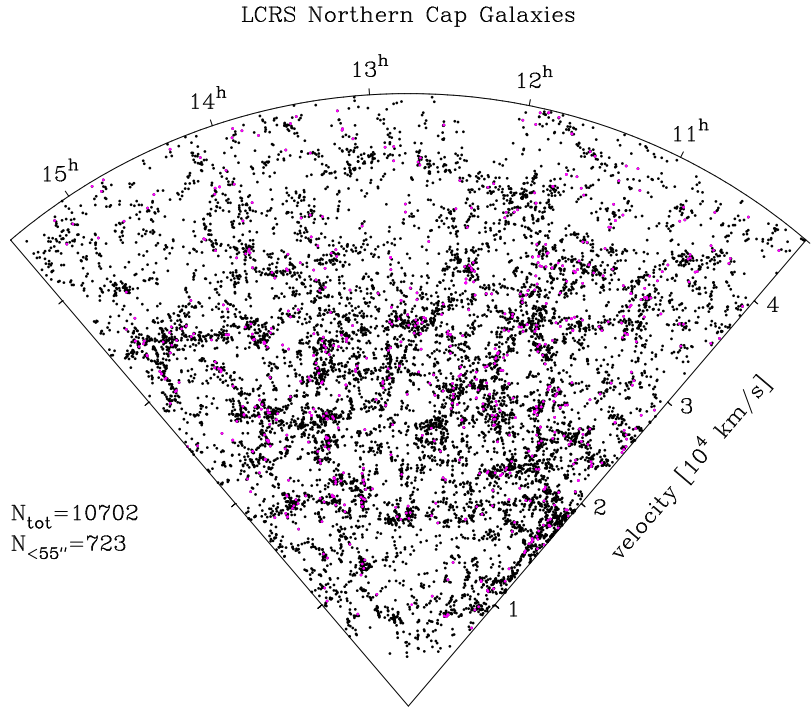}{5.00in}{-90}{60}{60}{-225}{425}
\plotfiddle{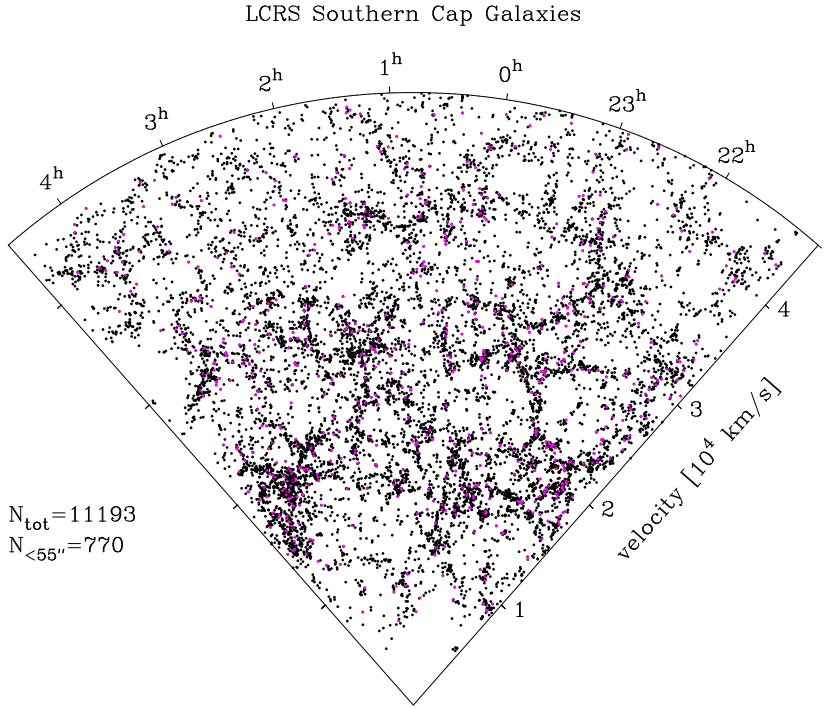}{5.00in}{-90}{60}{60}{-225}{500}
\vspace{-6.0cm}
\caption{The distribution of galaxies in the LCRS Northern (top) and
Southern (bottom) Galactics out to $cz = 46,000$~km~s$^{-1}$.  Only
those galaxies having luminosity $-22.5 \leq M_R - 5\log h < -17.5$
and lying within the LCRS official geometric and photometric
boundaries are plotted.  Red points are the 55-arcsec ``orphans,''
plotted with their ``faked'' velocities.  $N_{\rm tot}$ is the total
number of galaxies plotted, 55-arcsec ``orphans'' included; $N_{\rm
<55''}$ refers to the number of 55-arcsec ``orphans'' plotted.}
\end{figure}
 
\clearpage

\begin{figure}
\plotfiddle{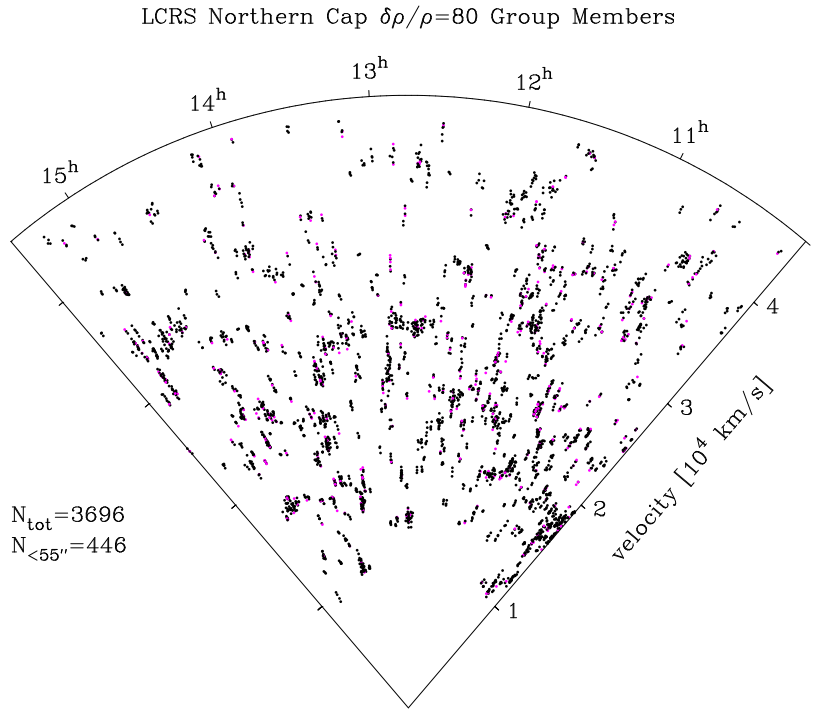}{5.00in}{-90}{60}{60}{-225}{425}
\plotfiddle{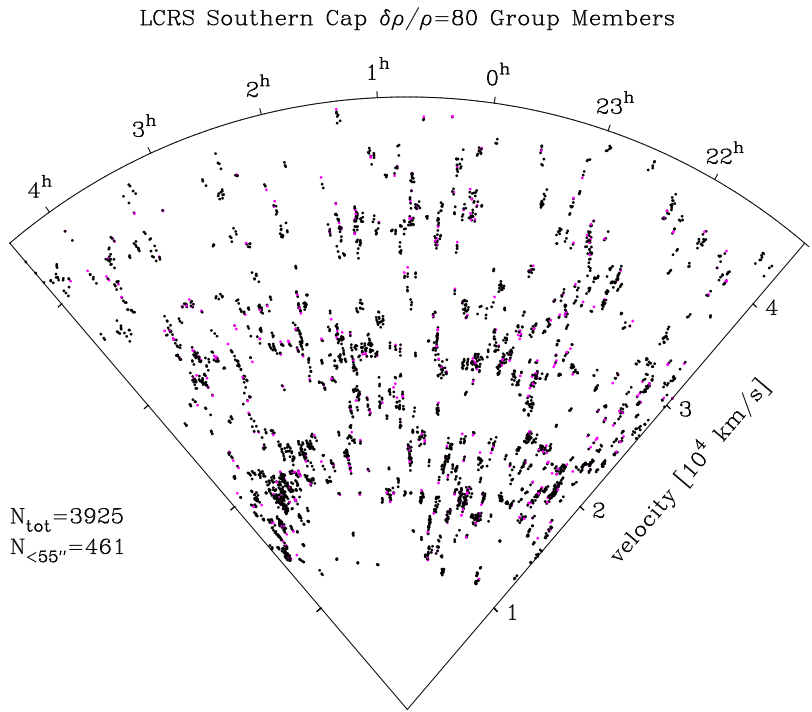}{5.00in}{-90}{60}{60}{-225}{500}
\vspace{-4.0cm}
\caption{Same as Figure~1, but only galaxies in $\delta\rho/\rho = 80$
groups are plotted.}
\end{figure}

\clearpage

\begin{figure}
\plotfiddle{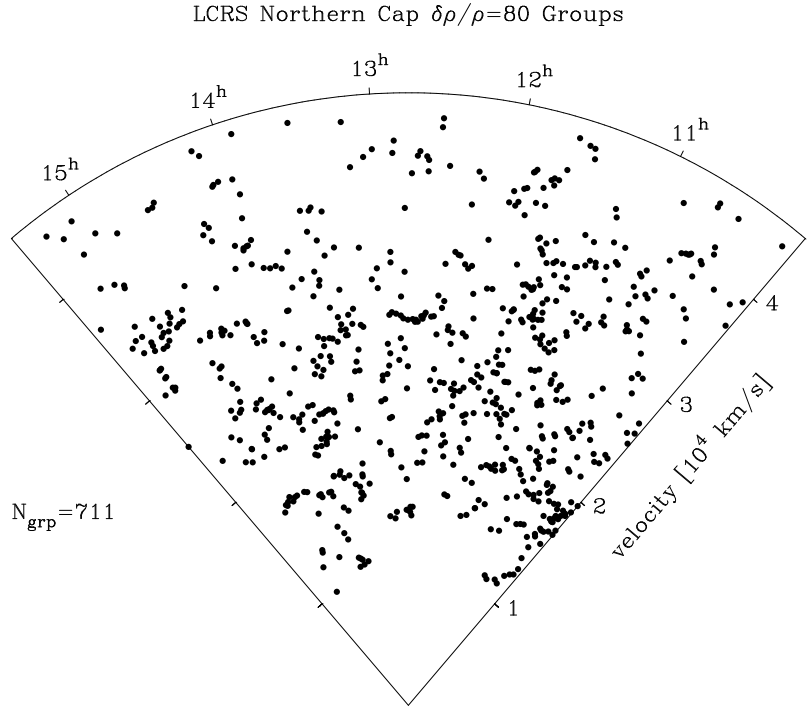}{5.00in}{-90}{60}{60}{-225}{425}
\plotfiddle{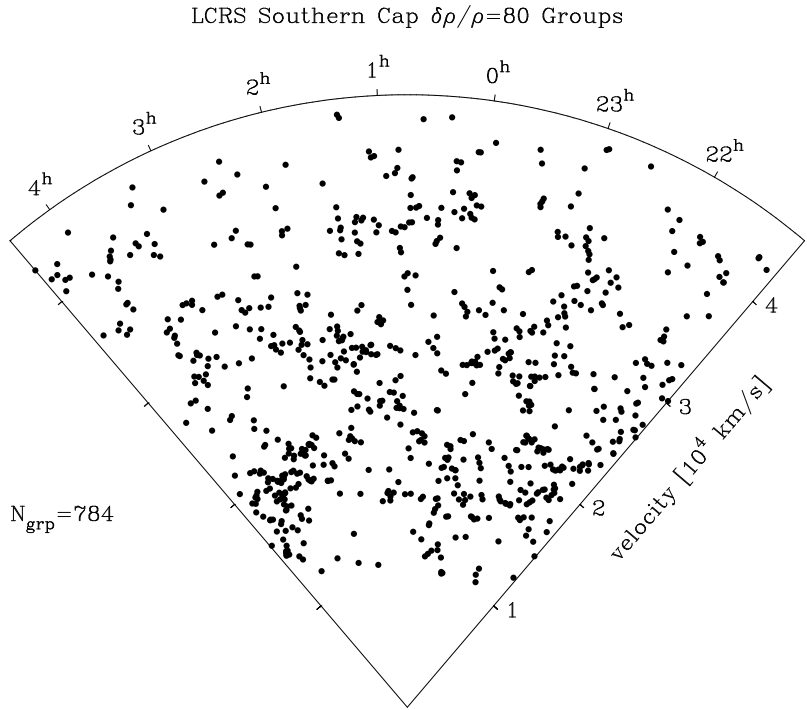}{5.00in}{-90}{60}{60}{-225}{500}
\vspace{-6.0cm}
\caption{The distribution of $\delta\rho/\rho = 80$ groups in the LCRS
Northern (top) and Southern (bottom) Galactic Caps.  The LCRS Group
Catalogue extends from $cz = 10,000$~km~s$^{-1}$ to $cz =
45,000$~km~s$^{-1}$; so the dearth of groups at $cz <
10,000$~km~s$^{-1}$ is not physical but merely the cutoff of the
catalogue.}
\end{figure}

\clearpage

\begin{figure}
\plotfiddle{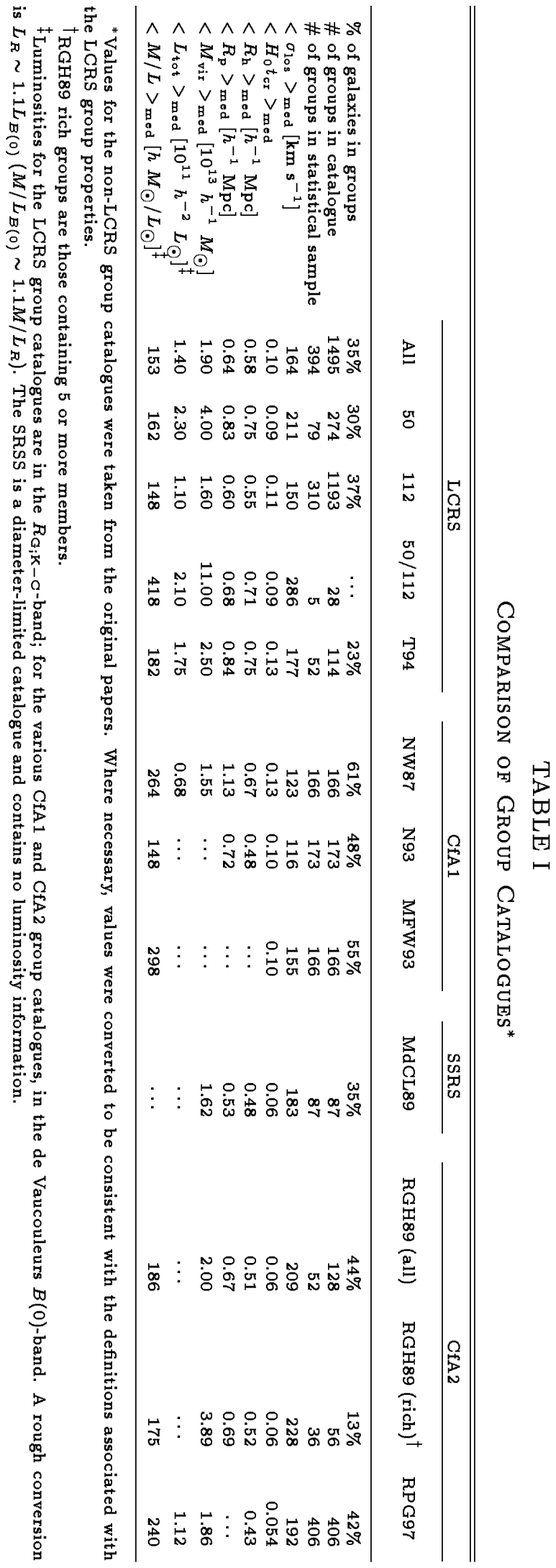}{5.00in}{0}{120}{120}{-225}{-290}
\end{figure}

\end{document}